*Article*

# Combining visual contrast information with sound can produce faster decisions

**Birgitta Dresp-Langley\*+, Marie Monfouga+**

⁺ Centre National de la Recherche Scientifique UMR 7357 ICube Lab, Strasbourg University, FRANCE;

\* Correspondence: birgitta.dresp@unistra.fr; Tel.: +33-388119117

**Abstract**

Pieron's and Chocholle's seminal psychophysical work predicts that human response time to information relative to visual contrast and/or sound frequency decreases when contrast intensity or sound frequency increases. The goal of this study is to bring to the fore the ability of individuals to use visual contrast intensity and sound frequency in combination for faster perceptual decisions of relative depth ("nearer") in planar (2D) object configurations on the basis of physical variations in luminance contrast. Computer controlled images with two abstract patterns of varying contrast intensity, one on the left and one on the right, preceded or not by a pure tone of varying frequency, were shown to healthy young humans in controlled experimental sequences. Their task (two-alternative forced-choice) was to decide as quickly as possible which of two patterns, the *left* or the *right* one, in a given image appeared to "stand out as if it were nearer" in terms of apparent (subjective) visual depth. The results show that the combinations of varying relative visual contrast with sounds of varying frequency exploited here produced an additive effect on choice response times in terms of facilitation, where a stronger visual contrast combined with a higher sound frequency produced shorter forced-choice response times. This new effect is predicted by cross-modal audio-visual probability summation.

**1. Introduction**

On the basis of predictions derived from Pieron's Law [1], classic psychophysical response times studies using two-alternative forced choice techniques have shown that human response time to contrast information decreases when the luminance intensity of a stimulus, or the contrast between two stimuli, increases [2]. Moreover, for a constant luminance or contrast level, response times decrease when the visual area of contrast increases because of a probability summation effect [3] in the contrast processing channels of the visual brain. Ahead of Piéron, psychophysicists like Exner, Wundt, Cattell, and Chocholle [4,5,6] had already emphasized the inverse relationship between human response time and stimulus intensity, in different sensory modalities including sound. Chocholle [7] and subsequently Stevens [8] systematically investigated human motor response time as a function of loudness (dB) and/or sound frequency (Hz), showing that an increase in either parameter may produce a decrease in response times, or the perceptual system's readiness to respond. Since these early and seminal psychophysical studies, further research has shown that sound information impacts on information processing by other senses including vision, and may considerably influence our decisions in response to signals we receive [9, 10].

The human brain's capacity to exploit combined information of visual contrast and sound in motor response behavior [11] has important implications in the context of a variety of operator tasks in the context of human-computer interaction systems where optimal motor performance is critical [12, 13, 14]. The goal of this study here was to bring to the fore the ability of individuals to use visual contrast and sound effectively for making faster perceptual decisions by taking into account the well documented capacity of the human perceptual system to extract subjective cues of relative depth from planar (2D) object configurations on the basis of physical variations in luminance contrast [15-27]. As shown previously, in 2D configurations with higher contrast and a lower contrast visual objects, those with the higher contrast will be consistently perceived as "nearer" by human observers. The greater the difference in contrast between two objects in a 2D image, the higher is the probability for the stronger contrast to be perceived as "nearer" [18] and, as a direct consequence of Piéron's Law [1], the shorter will be the time taken to reach that perceptual decision [18].

Combining visual contrast differences with pure sounds of varying frequency should produce summative effects on response times for "nearer" in this context under a probability summation hypothesis, where stronger contrasts combined with higher sound frequencies lead to faster perceptual decisions. This hypothesis was investigated taking into account that identical sounds, in terms of physical intensity (dB), with higher frequencies have higher average energy for any given section of the sound wave compared with lower frequency sounds. This results in the perception of differences in pitch [28], where sounds with a higher frequency are subjectively assimilated to sounds of a higher intensity [29] although physically they have the same intensity in dB.

**2. Materials and Methods**

Stimulus sequences (images and sounds) in the different experimental conditions, corresponding to individual trial sessions, and data acquisition were computer controlled. The experimental task was a classic psychophysical two-alternative forced choice (2AFC) task [30], yielding perceptual decisions relative to perceived relative pattern depth in this study here, and their associated decision times, more generally referred to as choice response times.

*Research ethics and participants*

The experiments were conducted in conformity with the Helsinki Declaration for scientific experiments on human individuals, and in full compliance with regulations set by the ethics board of the corresponding author's host institution (CNRS) relative to response data collection from healthy human individuals in non-invasive psychophysical tasks. Ten healthy young individuals, five men and five women, took part in the experiments as undergraduate study volunteers. All had normal vision and hearing, and provided informed consent to participate as subjects. Their identity is not revealed.

*Image and sound conditions*

Image configurations for the experiments were computer generated and displayed on a high resolution color monitor (EIZO COLOR EDGE CG 275W, 2560 x 1440) connected to a DELL T5810 computer (Intel Xeon CPU E5-1620), equipped with an NVidia GForce GTX980 graphics card and a sound card with port for plugging in headphones. Color and luminance calibration of the RGB channels of the monitor was performed using the inbuilt Color Navigator self-calibration software, which is delivered with the screen and runs under Windows 7. RGB values here correspond to ADOBE RGB. All luminance levels were cross-checked with an external photometer (OPTICAL, Cambridge Research Systems). RGB coordinates and luminance parameters (cd/m$^2$) of the different patterns in the image configurations and their dark and light backgrounds are given in Table 1.

Weber contrasts (*LumC*) in the different positive and negative polarity displays produced systematic differences in contrast (*dC*) between left and right patterns (Table 1) of an image pair. Within this range, *dC* are predicted to produce a high-probability (between .95 and 1) "foreground" effect in the stronger of the two pattern contrasts, as explained in the introduction. Patterns had variable number of elements across image pairs, but never within (see Figure 1). The size of each square surface in the patterns was 16x16 pixels, the size of a single pixel on the screen being 0.023 cm. Lighter and darker patterns were paired (Figure 1), and randomly displayed to the left and to the right in alternation. All configurations were displayed centrally on the monitor in computer controlled sequences on their dark or light backgrounds. Task sessions and data generation were controlled by a program written in Python for Windows and available online at:

https://pumpkinmarie.github.io/ExperimentalPictureSoftware/

**Table 1.** RGB values and luminance parameters (*Lum*) in *candela* per square meter (*cd*/m2) for patterns with positive (light on dark) and negative (dark on light) contrast sign (polarity). Lighter and darker patterns were paired in the image configurations (Figure 1) and displayed to the left and to the right. *LumC* corresponds to Weber contrasts, calculated as given in (1). The difference between the Weber contrasts (*dC*) of two patterns in a pair determines the perceived difference in relative pattern depth.

|  |  | R | G | B | *Lum* | LumC | dC |
|---|---|---|---|---|---|---|---|
| **Dark image background** |  | 5 | 5 | 5 | 2.5 (*cd*/m²) |  |  |
| **Light image background** |  | 250 | 250 | 250 | 95 (*cd*/m²) |  |  |
| **Positive-sign light-on-dark pairs** |  |  |  |  |  |  |  |
| '*dC* +' | lighter patterns | 250 | 250 | 250 | 95 (*cd*/m²) | 37 |  |
|  | darker patterns | 150 | 150 | 150 | 52 (*cd*/m²) | 20 | 17 |
| '*dC* ++' | lighter patterns | 250 | 250 | 250 | 95 (*cd*/m²) | 37 |  |
|  | darker patterns | 100 | 100 | 100 | 30 (*cd*/m²) | 11 | 26 |
| '*dC* +++' | lighter patterns | 250 | 250 | 250 | 95 (*cd*/m²) | 37 |  |
|  | darker patterns | 50 | 50 | 50 | 10 (*cd*/m²) | 3 | 34 |
| **Negative-sign dark-on-light pairs** |  |  |  |  |  |  |  |
| '*dC* -' | darker patterns | 5 | 5 | 5 | 2.5 (*cd*/m²) | 37 |  |
|  | lighter patterns | 50 | 50 | 50 | 10 (*cd*/m²) | 8.5 | 28.5 |
| '*dC* - -' | darker patterns | 5 | 5 | 5 | 2.5 (*cd*/m²) | 37 |  |
|  | lighter patterns | 100 | 100 | 100 | 30 (*cd*/m²) | 2.2 | 34.8 |
| '*dC* - - -' | darker patterns | 5 | 5 | 5 | 2.5 (*cd*/m²) | 37 |  |
|  | lighter patterns | 150 | 150 | 150 | 52 (*cd*/m²) | 0.8 | 36.2 |

Positive-sign and negative-sign pattern contrasts are expressed here in terms of Weber contrast (*LumC*), which is given by

$$LumC = (Lum\_max - Lum\_min) / Lum\_min \qquad (1).$$

The difference in visual contrast (*dC*) between two patterns in a pair is given by

$$dC = LumC\_max - LumC\_min \qquad (2).$$

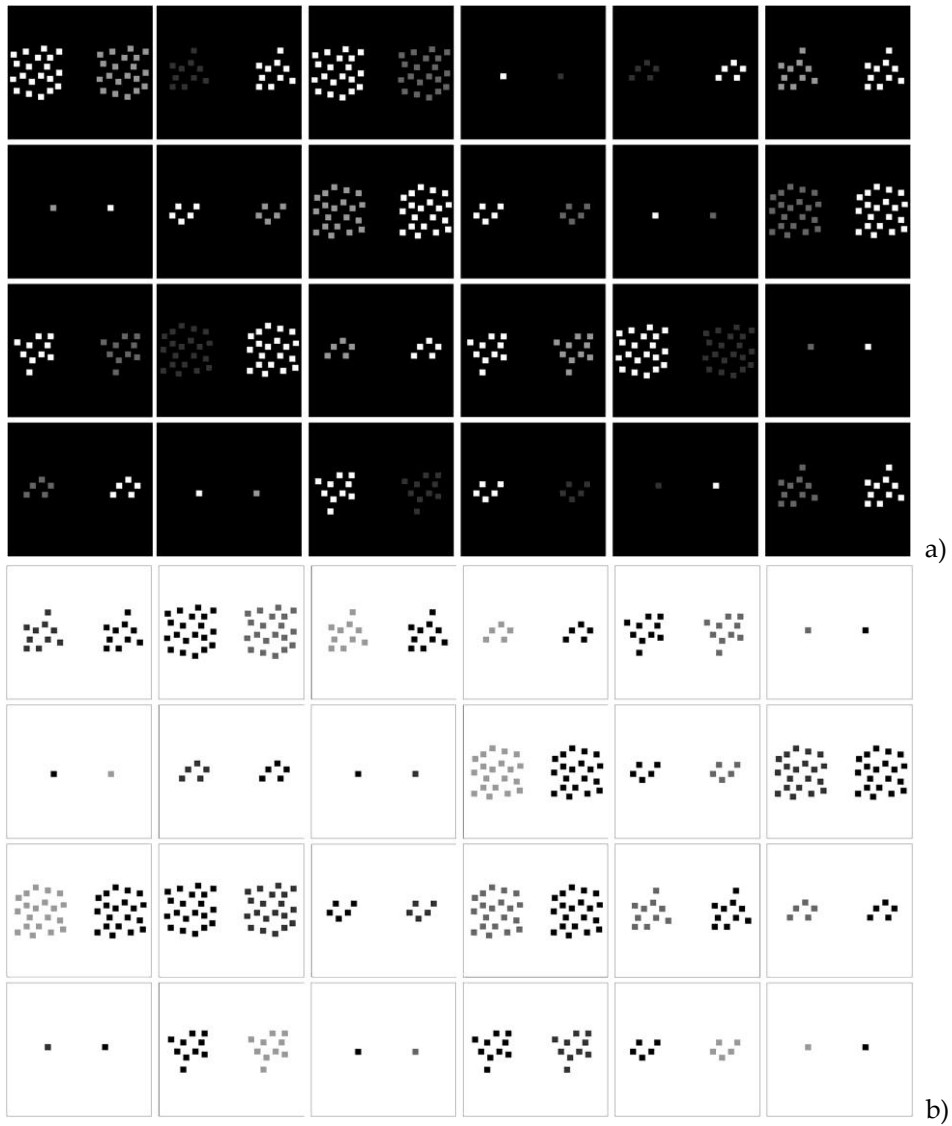

**Figure 1.** 48 paired image configurations with variable contrast intensities used as stimuli in the experiments. 24 pairs had positive (a) and 24 pairs negative (b) contrast polarity. Each such pair was preceded by a 70 dB pure tone sound signal (200, 1000, or 2000 Hz) in test conditions with sound.

Pure tone sound signals with three different sound frequencies, corresponding to 200, 1000 and 2000 hertz (Hz), with identical amplitude of 70 decibels (Db), were generated from a wav file. Sound frequency (Hz) measures the speed with which a sound wave propagates and determines the pitch of a sound. Human individuals with normal hearing are perfectly able to discriminate variations in pitch within an acoustic range between 20 Hz and 20 000 Hz. Within that range, higher pitch sounds are perceived as "sharper" than lower pitch sounds of the same amplitude. For illustration, sound waves of 200 Hz and 100 Hz with identical amplitude are displayed here below in Figure 2.

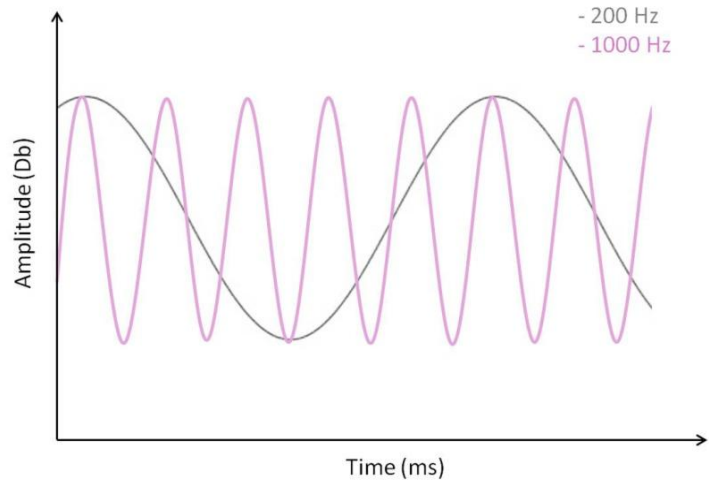

**Figure 2.** Illustration of a 200 Hz, and a 1000 Hz sound wave of identical amplitude (Db). Within the audible frequency range (20 Hz to 20 000 Hz) for humans, higher frequency (higher pitch) sounds of a given amplitude are perceived as "sharper" than lower frequency sounds of identical amplitude.

*Experimental design*

Pattern pairs of light-on-dark and dark-on-light contrast with varying number of pattern elements (Figure 1) were displayed in a random order in separate counterbalanced experimental sessions for each o the two conditions of the contrast polarity factor (*Polaritie$_2$*). The number of pattern elements (E) on both sides of a pair varied between n=1, n=5, n=10, and n=20 (see Figure 1 for illustration), yielding another factor of systematic variation with four levels (*Elements$_4$*). The contrast intensity of patterns in image pairs varied in such a way that the strongest pattern contrast (see Table 1) was always associated with a weaker pattern contrast of the same polarity, and presented to the left and to the right in alternation in a given image pair. This produced three levels of difference in pattern contrast (*dC*), within and across polarity conditions, yielding a factor of systematic variation with three levels (*dC$_3$*), and a secondary factor of relative location with two levels (*Locations$_2$*), not expected to produce any systematic effects on perceptual responses. Each image pair was preceded by a 100 millisecond (ms) pure tone sound signal with a frequency of either 200 Hz, 1000 Hz, 2000 Hz, or 0 Hz ('no sound' control condition), yielding another factor of systematic variation with four levels (*Sound$_4$*). The delay between the end of a given sound signal and the beginning of a given image presentation on each single trial was 800 milliseconds. Different sound conditions were presented in separate counterbalanced experimental sessions. With ten individuals (*Individual$_{10}$*) run in separate trial block sessions, we have the following experimental design plan: *Individuals$_{10}$* x *Polarities$_2$* x *dC$_3$* x *Locations$_2$* x *Elements$_4$* x *Sounds$_4$*, producing a total number of N = 1920 experimental observations, with 192 data per subject, in terms of response times and their associated perceptual decisions.

*Procedure and task instructions*

The subject was comfortably seated in front of the computer at a distance of about 80 cm from the screen in a semi-dark room (mesopic viewing condition) and adapted to surrounding conditions for about five minutes. He/she was informed that images with two abstract patterns, one on the left and one on the right, will be shown in sequences, preceded or not by a brief tone, and that his/her task was to decide as quickly as possible which of the two patterns, the *left* or the *right* one, in a given image appeared to "stand out as if it were nearer" in terms of apparent (subjective) visual depth, as previously in [17, 18, 21, 23, 24]. A response had to be delivered by pressing '1' for '*left*' or '2' for '*right*', and was recorded and stored in a labeled data column of an excel file. The response time, i.e. the time between an image onset and the moment a response key was pressed, was also recorded by the computer, and stored in a second labeled data column of the same excel file. As soon as a response was given, the image disappeared from the screen, and 900 milliseconds later the next image of a given sequence appeared. In the conditions where the images were preceded by a 100 ms sound signal of a given frequency, the sound was delivered after 800 milliseconds following the previous response.

## 3. Results

The choice response time data and their associated perceptual decisions ('nearer on left' *versus* 'nearer on right') were analyzed to evaluate combined effects of visual contrast information and sound frequency, i.e. pitch, on the time taken to make a perceptual decision.

*Perceptual decisions relative to expected depth effects ("nearer")*

As explained in the introduction, such an analysis makes sense provided the perceptual decisions are 'high-probability', i.e. reflect very little stimulus uncertainty. To meet this requirement, the contrast differences between patterns of a pair in the images here were chosen in the light of previous studies [17, 18], under the prediction that they would produce high-probability effects of perceived relative depth reflected by a 95% to 100% decision rate for "nearer" in response to the stronger contrast patterns of the pairs. This prediction was confirmed. For the 24 positive contrast polarity images, a 98% response rate for "nearer" to the stronger contrast pattern in a pair was recorded, and for the 24 negative contrast polarity images, we have a 96% response rate for "nearer" to the stronger contrast pattern of a pair.

*Effects of experimental factors on response times*

Response time data were analyzed in terms of means and standard errors for a graphical representation, shown here below in Figure 3, of effects of the different experimental factors. The individual response time data were fed into a Four-Way ANOVA (Analysis Of Variance) to assess the statistical significance of these effects. The analysis plan corresponds to the experimental design plan *Individuals*$_{10}$ x *Polarities*$_2$ x *dC*$_3$ x *Locations*$_2$ x *Elements*$_4$ x *Sounds*$_4$ with a total number of 1920 data points for individual response times. The source of random variability is the subject factor *Individuals*$_{10}$. The two levels of the secondary factor *Locations*$_2$, relative to counterbalanced variations in the spatial location of stronger/weaker patterns in a pair (left or right), are not associated with any hypothesis and, as expected, did not produce a noticeable difference in response times, as revealed by comparison between the means for these two secondary factor levels. The results of the ANOVA yielding statistically significant effects are summarized here below in Table 2, which shows the F statistics relative to effects, and their respective probability limits. The full set of raw data (individual response times) from which the analyses here are drawn is provided in Table S1 of the Supplementary Materials Section.

*Contrast polarity*

Effects of the polarity of pattern contrast on response times are shown here when comparing the graphs on the left of Figure 3 to the graphs on the right of Figure 3. Positively signed light-on-dark pattern pairs (Figure 3, graphs on left) produced shorter response times in comparison with negatively signed dark-on-light pattern pairs (Figure 3, graphs on right) despite the fact that the pattern pairs with negative contrast sign displayed moderately stronger differences in visual contrast (*dC*) between patterns in a pair. This effect of contrast polarity is statistically significant

(Table 2). It is explained by the well-documented functional asymmetry between the so-called "on" and "off" contrast processing channels in the human brain [18, 20, 27, 31]. One of the perceptual consequences of this functional asymmetry is that positively signed contrast configurations, processed by the "on" channels of the visual brain, produce stronger effects of figure-ground segregation [24] and relative depth [17], with shorter perceptual decision times, as confirmed by this result here.

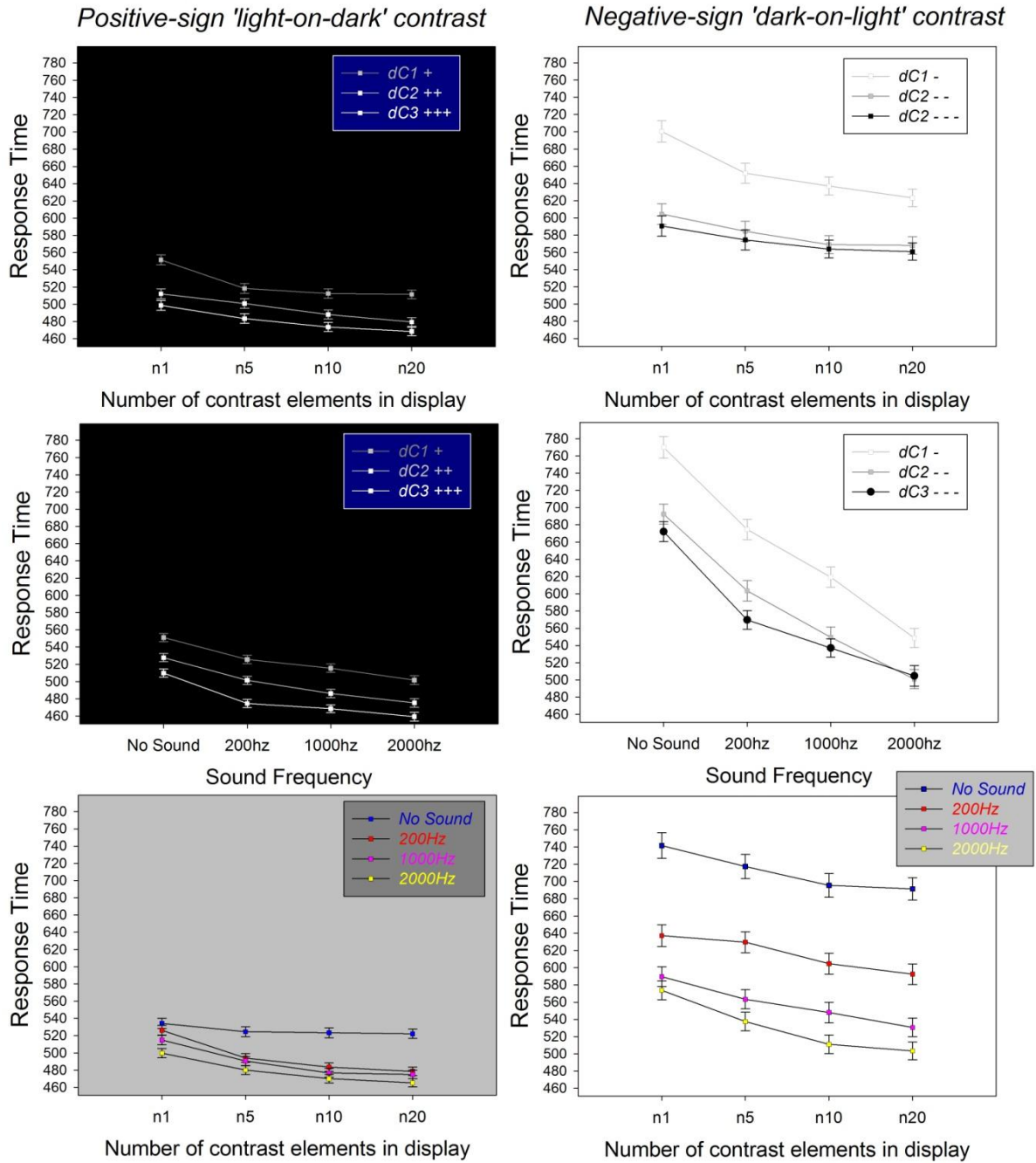

**Figure 3.** Graphic representation of the effects of relative visual contrast between patterns in a pair (*dC*), contrast sign, number of contrast elements, and sound on perceptual decision times from this study. Mean response times and their standard errors are plotted to show effect sizes and interactions.

*Contrast difference (dC) in a pattern pair*

Effects of the difference in visual Weber contrast ($dC$) between two patterns of a pair on response times are displayed in the two graphs on top as a function of contrast sign and number of contrast elements, and in the two graphs in the middle of Figure 3 as a function of contrast sign and sound frequency. These plots show that response times consistently decrease as the $dC$ increases, in pattern pairs with positive contrast sign (Figure 3, top and middle left) and in pattern pairs with negative contrast sign (Figure 3, top and middle right). This effect of $dC$ on response times reflecting perceptual decisions for relative depth ("nearer") is statistically significant (Table 2), and predicted by results from previous studies [17, 18], as explained in the introduction, and summarized further here below in the discussion.

*Number of contrast elements in a pattern pair*

Effects of the number of contrast elements in a pattern pair on perceptual response times for "nearer" are displayed in the two graphs on top of Figure 3 as a function of contrast sign and number of contrast elements, and in the two graphs at the bottom of Figure 3 as a function of contrast polarity and sound frequency. These plots show that response times consistently decrease as the number of contrast elements in the patterns increases, in pattern pairs with positive contrast sign (Figure 3, top and bottom left) and in pattern pairs with negative contrast sign (Figure 3, top and bottom right). This effect of the number of contrast elements in the patterns on response times is also statistically significant (Table 2), and is explained by spatial probability summation in the "on" and "off" contrast processing channels of the visual brain, as pointed out again further below in the discussion.

*Sound Frequency*

Effects of sound frequency on perceptual response times for "nearer" are displayed in the two graphs in the middle of Figure 3 as a function of contrast sign, and in the two graphs at the bottom of Figure 3 as a function of the number of contrast elements. These plots show that response times consistently decrease as the sound frequency increases, in pattern pairs with positive contrast sign (Figure 3, middle and bottom left) and in pattern pairs with negative contrast sign (Figure 3, middle and bottom right). The effect of sound frequency on response times is statistically significant (Table 2).

**Table 2.** Results from the 4-Way ANOVA on the response time data (N = 1920) with **F** statistics relative to effects of factors and their interactions, degrees of freedom (*df*) of the given comparison, and statistical probability limits (*p*).

| Factor | df | F | p |
|---|---|---|---|
| Polarity | 1 | 231.926 | <.001 |
| Nelements | 3 | 3.397 | <.017 |
| dC | 2 | 24.990 | <.001 |
| Sound Frequency | 3 | 49.835 | <.001 |
| **Interactions** | | | |
| Nelements x dC | 6 | 0.872 | .515 NS |
| Nelements x Sound Frequency | 9 | 0.307 | .973 NS |
| dC x Sound Frequency | 6 | 0.727 | .628 NS |
| Nelements x Polarity | 3 | 0.845 | .535 NS |
| dC x Polarity | 2 | 3.891 | <.021 |
| Sound Frequency x Polarity | 3 | 20.880 | <.001 |

*Interactions*

Possible interaction between effects of the factors tested here are shown graphically in Figure 3. There is no significant interaction between the number of contrast elements (*N*elements) and any of the other three factors (Table 2), nor is there a significant interaction between the sound frequency and the difference in visual contrast (*dC*) of patterns in a pair (Table 2). Interactions between *dC* and contrast polarity, and between sound frequency and contrast polarity are statistically significant (Table 2). *Post-hoc* paired comparisons (Holm-Sidak tests) were computed for factor levels relative to the significant interactions to unravel which paired comparisons between factor levels yield statistical significance. The results from these analyses are provided in Table S2 of the Supplementary Materials Section.

## 4. Discussion

As predicted by probability summation [1, 2, 3, 4, 7, 8], combinations of visual contrast and sounds of varying frequency should produce additive effects on choice response times. This prediction is confirmed by the results of the experiments here. Variations in luminance contrast were used to manipulate relative depth in 2D images producing perceptual decisions for "nearer" [17, 18]. It is shown that stronger contrasts combined with higher sound frequencies lead to faster perceptual decisions [17, 18]. This facilitating effect of sound frequency on response times for "nearer" was consistently stronger in the positively signed, light-on-dark, contrast configurations, as predicted by functional asymmetries between the "on" and "off" contrast processing channels of the visual brain [18, 21, 27, 31]. Moreover, as the number of contrast elements in the 2D patterns increases, the effect of sound on response times also increases statistically, regardless of the contrast sign of the patterns, as predicted by spatial probability summation in the "on" and "off" contrast processing channels of the visual brain. There is no interaction between number of contrast elements in the patterns and their contrast polarity. These results lead to conclude that sound frequencies can be effectively used to produce faster decisions in specific visual tasks where the processing of contrast information is critical. The pure tone sound signals preceding the visual contrast stimuli here had three different sound frequencies and identical amplitude, generated to manipulate the speed with which the sound wave propagates and determines the perceived pitch of each sound. Within the audible frequency range, higher pitch sounds are generally perceived as "sharper" or "louder" than lower pitch sounds of the same amplitude. After the experimental trials here, all subjects in the post-test debriefing stated having perceived some of the tones as considerably "sharper" or "louder" than others. In terms of the effect of the different tones on the times taken to reach perceptual decisions for "nearer", the 2000 Hz tones with the most wave energy, potentially yielding the highest pitch, consistently produced the strongest facilitation effects on response times compared with the no-sound control condition.

The human brain has to analyze and react in real time to an enormous amount of information from the eyes, ears and other senses. How all this information is efficiently represented and processed in the nervous system is a complex topic in nonlinear and complex systems research. It has been suggested that dynamical attractors may form the basis of all neural information processing [24, 28, 29, 31]. The auditory and visual systems are, indeed, complex and highly nonlinear physiological systems. The combined processing of information from different sensory channels carries perceptual and functional meaning, as highlighted by the results from this study here.

**Supplementary Materials:** The following are available online at www.mdpi.com/link, Table S1: Raw data (individual response times) for the different experimental conditions as fed into the 4-Way

ANOVA Table S2: Results of *the post-hoc* paired comparisons (Holm-Sidak tests) between factor levels relative to significant interactions.

**Funding:** This research received no external funding.

**Acknowledgments:** The authors are grateful to the students who volunteered as participants, and provided post-test insights with comments and suggestions for future experiments.

**Conflicts of Interest:** The author declares no conflict of interest.

**References**

1. Piéron, H. *The Sensations.* 1952; New Haven, CT: Yale University Press.
2. Pins, D, Bonnet, C. On the relation between stimulus intensity and processing time: Pieron's law and choice reaction time. Perception & Psychophysics. 1996; 58: 390–400.
3. Bonnet, C, Gurlekian, J, Harris, P. Reaction time and visual area: Searching for the determinants *Bulletin of the Psychonomic Society*. *1992; 30* (5): 396-398.
4. Cattell, J. M. The influence of the intensity of the stimulus on the length of the reaction time. *Brain.* 1886; 8: 512-515.
5. Exner, S. Ueber die zu einer Gesichtswahrnehmung noetige Zeit. *Sitzungsberichte der Kaiserlichen Akademie der Wissenschaften,* 1868; 57: 601-632.
6. Wundt, W. *Grundzüge der Physiologischen Psychologie.* 1874 ; Leipzig: Engelmann.
7. Chocholle. R. (1940). Variation des temps de réaction auditifs en fonction de l'intensité a diverses fréquences. *L'Année Psychologique.* 1940; 41/42 :65-124.
8. Stevens, SS. *Psychophysics: Introduction to its perceptual neural and social prospects.* 1975; New York: Wiley.
9. Altmann J. Acoustic weapons: a prospective assessment. 2001; *Sci Glob Secur* 9: 165–234. Doi: 10.1080/08929880108426495
10. Beckerman J. The sonic boom. how sound transforms the way we think, feel, and buy. 2014; Boston, New York: Houghton Mifflin Harcourt.
11. Hanes, D, Schall, J. Neural control of voluntary movement initiation. *Science.* 1996; 274: 427–430
12. Batmaz AU, de Mathelin M, Dresp-Langley B. Getting nowhere fast: Trade-off between speed and precision in training to execute image-guided hand-tool movements. *BMC Psychology.* 2016a; 4: 55. doi:10.1186/s40359-016-0161-0.
13. Batmaz, AU, de Mathelin, M, Dresp-Langley, B. Seeing virtual while acting real: Visual display and strategy effects on the time and precision of eye-hand coordination. *PLoS One.* 2017; 12: e0183789.
14. Dresp-Langley, B. Principles of perceptual grouping: implications for image-guided surgery, *Frontiers in Psychology*, 2015; 6: 1565.
15. Farnè, M. Brightness as an indicator to distance: relative brightness *per se* or contrast with the background? *Perception.* 1977; 6: 287-293.
16. Egusa, H. Effects of brightness, hue, and saturation on the perceived depth between adjacent regions in the visual field. *Perception.* 1983; 12: 167-175.
17. Dresp, B, Durand, S, Grossberg, S. Depth perception from pairs of overlapping cues in pictorial displays, *Spatial Vision.* 2002; 15: 255–276.
18. Guibal C, Dresp, B. Interaction of color and geometric cues in depth perception: When does "red" mean "near"?" *Psychological Research*, 2004; 10: 167-178.
19. Qiu, T. Sugihara, von der Heydt, R. Figure-ground mechanisms provide structure for selective attention. *Nature Neuroscience.* 2007; 11, 1492-9.
20. Dresp, B, Fischer, S. Asymmetrical contrast effects induced by luminance and colour configurations. *Perception & Psychophysics.* 2001; 63: 1262-1270.
21. Dresp-Langley, B, Reeves, A. Simultaneous brightness and apparent depth from true colors on grey: Chevreul revisited. *Seeing and Perceiving.* 2012; 25: 597-618.

22. Dresp-Langley, B, Reeves, A. Color and figure-ground: From signals to qualia. In S. Magnussen, M. Greenlee, J. Werner, A. Geremek (Eds.): *Perception beyond Gestalt: Progress in Vision Research*. 2014; Psychology Press, Abingdon (UK), pp. 159-71.
23. Dresp-Langley, B., Reeves, A. Effects of saturation and contrast polarity on the figure-ground organization of color on gray. *Frontiers in Psychology*, 2004; 5: 1136.
24. Dresp-Langley, B, Grossberg, S. Neural Computation of Surface Border Ownership and Relative Surface Depth from Ambiguous Contrast Inputs, *Frontiers in Psychology*. 2016; 7: 1102.
25. von der Heydt, R. Figure–ground and the emergence of proto-objects in the visual cortex, *Frontiers in Psychology.* 2015; 6: 1695.
26. *Dresp-Langley*, B, *Reeves*, A. Colour for behavioural success. *i-Perception*. 2018; 9(2): 1–23.
27. Spillmann, L, Dresp-Langley, B, Tseng, CH. Beyond the classic receptive field: The effect of contextual stimuli. *Journal of Vision*. 2015 15: 7.
28. Oxenham AJ. Pitch perception. *J Neurosci*. 2012;32(39):13335–13338. doi:10.1523/JNEUROSCI.3815-12.2012
29. Patel, A., Balaban, E. Human pitch perception is reflected in the timing of stimulus-related cortical activity. *Nature.* 2001; 4(8): 839-844.
30. Green, DM, Swets, JA. *Signal detection theory and psychophysics*. 1973; Krieger Publishing, Huntington, NY.
31. Schiller PH, Sandell JH, Maunsell JH. Functions of the ON and OFF channels of the visual system. *Nature.* 1986; 322(6082):824-5.